\documentstyle[toto,numreferences,epsfig]{balkan}

\begin{opening}
\title{DARK MATTER IN THE COSMOS-}
\subtitle{Its Direct Detection and the role of Nuclear Physics.}

\author{
J.D. Vergados$^{a,b}$}

\institute{
$~^A$Theoretical Physics Division, University of Ioannina, Gr 451 10,
Ioannina, Greece.\\
$^B$T-DO, Theoretical Physics Division, LANL, Los Alamos, N.M. 87545.}

\end{opening}

\runningtitle{DARK MATTER IN THE COSMOS}

\begin{document}
\bibliographystyle{plain}
\newcommand{\newc}{\newcommand}
\newc{\ra}{\rightarrow}
\newc{\lra}{\leftrightarrow}
\newc{\beq}{\begin{equation}}
\newc{\eeq}{\end{equation}}
\newc{\barr}{\begin{eqnarray}}
\newc{\earr}{\end{eqnarray}}


{Exotic dark matter together with dark energy or  cosmological constant
seem to dominate in the Universe. An even higher
density of such matter seems to be gravitationally trapped in our Galaxy.
The nature of dark matter can be unveiled only, if it is detected in the laboratory. 
Thus the accomplishment of this task
is central to physics and cosmology. Current fashionable supersymmetric models provide
 a natural dark matter candidate, which is the lightest
supersymmetric particle (LSP). Such models combined with fairly well
understood physics like
the quark substructure of the nucleon can yield the effective transition
operator at the nucleon level. We know, however, that the LSP is much heavier than the proton,
while its average energy is in the keV region. Thus the most likely possibility for its
direct detection is via its elastic scattering with a nuclear target by  observing
 the recoiling nucleus. In order to evaluate the event rate
one needs the  nuclear structure (form factor
and/or spin response function) for the special nuclear targets of experimental
interest.
 Since the expected rates for neutralino-nucleus scattering are expected to
 be small, one should exploit
 all the  characteristic signatures of this reaction. Such are: (i) In the standard recoil
measurements  the modulation of the event rate due to the Earth's motion.
(ii) In directional recoil experiments  the correlation of the event
 rate with the sun's motion. One now has both modulation, which is much larger and
depends not only on time, but
on the direction of observation as well, and  a large
forward-backward asymmetry. (iii) In non recoil experiments   
gamma rays following the decay of excited states populated during the
Nucleus-LSP collision. Branching  ratios of about 6 percent are possible. (iv) novel
experiments in which one observes the electrons prod used during the collision
of the LSP with the nucleus. Branching ratios of about 10 per cent are possible.} 
\date{\today}

\section{Introduction}
One of the greatest discoveries of the last couple of decays is the realization
that the matter, which  can shine, constitutes a very small fraction of the universe.
The universe is composed primarily of dark matter and dark energy, which are incapable of
emitting light. If one defines $\Omega_i=\frac{\rho_i}{\rho_{c}}$ for the species $i$, where $\rho_c$
the critical density (see sec. \ref{cosmos}), the recent cosmological observations
 \cite{ALLEXP}, when combined, lead to:
$$\Omega_b=0.05, \Omega _{CDM}= 0.30, \Omega_{\Lambda}= 0.65$$
for ordinary matter, dark matter and dark energy respectively. Their sum is \cite{ALLEXP}
 $\Omega=1.11\pm 0.07$, i.e. unity within the experimental errors , which means 
that the universe is flat.

\begin{figure}
\begin{center}
\includegraphics[height=0.48\textheight]{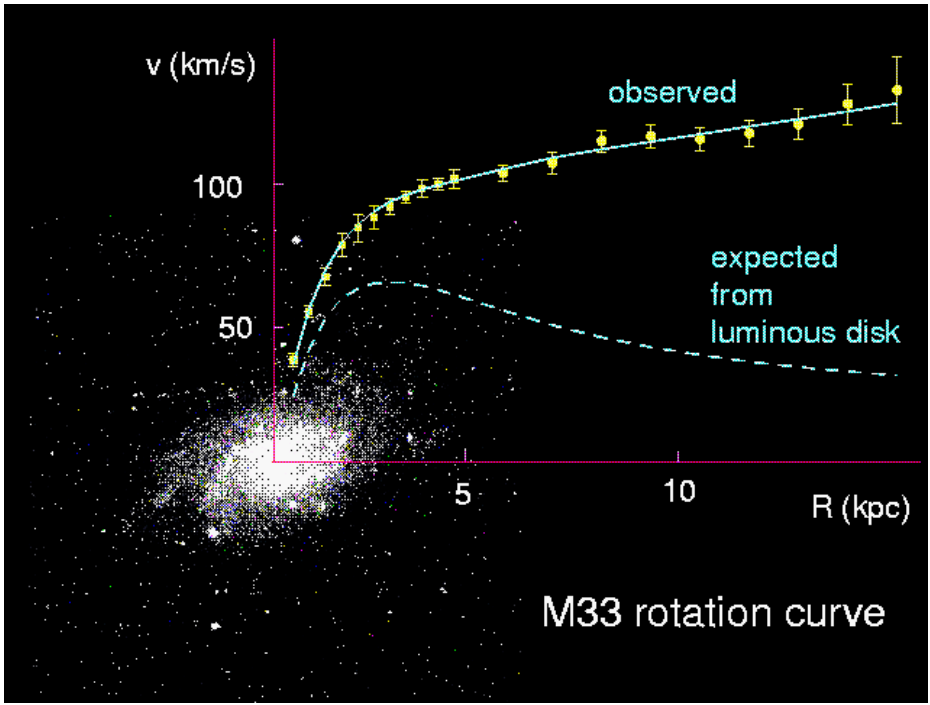}
\caption{The rotational velocity  distance. Clearly it does not fall outside the luminous matter.
This is suggestive of the presence of a dark matter halo outside the M33 galaxy.
\label{halos} }
\end{center}
\end{figure}

Additional indirect information
comes from  the rotational curves \cite{Jungm}. The rotational velocity of an object increases so long
is surrounded but matter. Once outside matter the velocity of rotation drops as the square root  of the 
distance. Such observations are not possible in our own galaxy. The observations of other galaxies, 
similar to our own, indicate that the rotational velocities of objects outside the luminous matter
do not drop (see Fig. \ref{halos}). So there must be a halo of dark matter out there.

  As we have already mentioned, however, it is only the  direct detection of dark matter in the
 laboratory, which  will
 unravel the nature of the constituents of dark matter.
In fact one such experiment, the DAMA,  has claimed the observation of such signals,
 which with better statistics has subsequently
been interpreted as  modulation signals \cite{BERNA}.
These  data, however, if they are due to the coherent process,
are not consistent with other recent experiments, see e.g. EDELWEISS and CDMS
\cite{EDELWEISS}.\\
 Supersymmetry naturally provides candidates for the dark matter constituents.
 In the most favored scenario of supersymmetry the
LSP can be simply described as a Majorana fermion (LSP or neutralino), a linear
combination of the neutral components of the gauginos and
higgsinos \cite{ref2}-\cite{ARNDU}. We are not going to address issues 
related to SUSY in this paper. Most SUSY models predict nucleon cross sections
much smaller the the present experimental limit
 $\sigma_S \le 10^{-5} pb$ for the coherent process. As we shall see below
the  constraint on the spin cross-sections is less
stringent.
 
  Since the neutralino is expected to be
 non relativistic with average kinetic energy $<T> \approx
40KeV (m_{\chi}/ 100 GeV)$, it can be directly detected
 mainly via the recoiling of a nucleus
(A,Z) in elastic scattering. 

In some rare instances the low lying excited states
may also be populated \cite{VQS03}. In this case one may observe
 the emitted $\gamma$ rays. Finally it has also recently been suggested \cite{VEJ04} 
that one may
be able to detect the neutralino by observing the ionization electrons produced
during its collision with the nucleus.

%
%

In every case to extract from the data information about SUSY from the 
relevant nucleon cross section, one must
know the relevant nuclear matrix elements
\cite{Ress}-\cite{DIVA00}. The static spin matrix elements used in the present
 work have been obtained from the literature \cite{VQS03}.

Anyway since in all types of experiments the obtained rates are very low, one would like to be able
to exploit the modulation of the event rates due to the earth's
revolution around the sun \cite{DFS86},\cite{Verg}. In order to accomplish this one
adopts a folding procedure, i.e one has to assume some velocity
distribution
\cite{DFS86,Verg,COLLAR92,GREEN04}
for the LSP. One also would like to exploit other signatures
expected to show up in directional experiments \cite{DRIFT}.
 This is possible, 
since the sun is moving with
relatively high velocity with respect to the center of the galaxy.

\section{Matter Density Versus Cosmological Constant (Dark Energy).}
\label{cosmos}

As we have mentioned in the previous section the universe is primarily made of
dark matter and dark energy. The nature of each one of them is not presently known.
We hope the future experiments to shed light on this.
The future of the universe depends on the composition and the nature of these two
constituents. In the present case we will briefly explore the possibility that the
dark energy density is constant, independent of the size of the universe, i.e. it
coincides with Einstein's cosmological constant.
                                                                                
 The evolution of the Universe is governed by the General Theory of
Relativity. The most
  commonly used  model is that of Friedman, which utilizes the Robertson-
Walker metric
\begin{equation}
 (ds)^2=(dt)^2-a^2(t)[\frac{(dr)^2}{1-\kappa r^2}
          +r^2((d \theta)^2+ \sin ^2 \theta (d \phi)^2)]
\label{cosmo.1}
\end{equation}
, where $a(t)$ is the scale factor. The resulting Einstein equations are:
\begin{equation}
 R_{\mu \nu}-\frac{1}{2}g_{\mu \nu}R=-8\pi G_{N}
T_{\mu \nu}+
                                      \Lambda g_{\mu \nu}
\label{kosmo.2}
\end{equation}
where $G_{N}$ is Newton's constant and $\Lambda$ is the
cosmological constant.
The equation for the scale factor $a(t)$ becomes:
\begin{equation}
\frac{d^{2}a}{dt^{2}}=-\frac{4\pi}{3}G_{N}(\rho+3p)a=-\frac{4\pi
                   G_{N}\rho}{3}a+\frac{\Lambda}{3}
\label{cosmo.3}
\end{equation}
where $\rho$ is the mass density. Then the energy is
\beq
E=\frac{1}{2}m(\frac{ar}{dt})^{2}-G_{N}\frac{m}{a}(4\pi\rho
a^{3})+\frac{\Lambda}{6}ma^{2}=constant=-\frac{\kappa}{2}m
\label{kosmo.4}
\eeq
 This can be equivalently be written as
\beq
 H^{2}+\frac{\kappa}{a^{2}}=\frac{8\pi}{3}G_{N}\rho+\frac{\Lambda}{3}
\label{cosmo.5a}
\eeq
 where the quantity $H$ is Hubble's constant defined by
\beq
 H=\frac{1}{a}\frac{da}{dt}
\label{cosmo.5b}
\eeq
Hubble's constant is perhaps the most important parameter of cosmology. In
fact it is not a constant but it changes with time. Its present day value
is given by
\beq
H_0 =( 65\pm 15)~km/s~M_{pc}^{-1}
\label{cosmo5.c}
\eeq
In other words $H_0^{-1}=(1.50\pm).35)\times10^{10}~y$, which is roughly equal
to the age of the Universe. Astrophysicists conventionally write it as
\beq
H_0 =100~h~km/s~M_{pc}^{-1}~~~~,~~~~~0.5<h<0.8
\label{cosmo5.d}
\eeq
 Equations \ref{cosmo.3}-\ref{cosmo.5a} coincide with those of the
Newtonian
  theory with the following two types of forces:
  An attractive force decreasing in absolute value with the scale
  factor (Newton)
  and a repulsive force increasing with the scale factor (Einstein):
                                                                                
\beq
  F=-G_{N} \frac{mM}{a^{2}}~~(Newton)~~,~~
  F=\frac{1}{3}\Lambda ma~~(Einstein)
\label{cosmo.6b}
\eeq
   Historically the cosmological constant was introduced by
   Einstein so that General Relativity yields a stationary
   Universe, i.e. one which satisfies the conditions:
\beq
    \frac{da}{dt}=0~~~~\frac{d^{2}a}{dt^{2}}=0
\label{cosmo.7}
\eeq
    Indeed for $\kappa>0$, the above equations lead to
    $a=a_{c}$=constant provided that
\beq
    \frac{1}{3}\Lambda a_{c}-\frac{4\pi}{3} G_{N}\rho a{c}=0~~,~~
     \frac{1}{3}\Lambda a^{2}_{c}-\frac{4\pi}{3} G_{N}\rho a^{2}_{c}=\kappa
\label{cosmo.8}
\eeq
    These equations have a non trivial solution provided that the density
$\rho$ and the cosmological constant $\Lambda$ are related, i.e.
\beq
    \Lambda=4 \pi G_{N} \rho
\label{cosmo.9}
\eeq
    The radius of the Universe then is given by
\beq
   a_{c}=[\frac{\kappa}{4\pi G_{N}\rho}]^{1/2}
\label{cosmo.10}
\eeq
 In our days we do not believe that the universe is static. So the two parameters,
 density and cosmological
constant, need not be related. Define now
\beq
\Omega_{m}=\frac{\rho}{\rho _{c}}~~,~~
          \Omega_{\Lambda}=\frac{\rho _{v}} {\rho _{c}}~~,~~
    \rho_{v}=\frac{\Lambda}{8 \pi G_{N}}~~("vacuum" density)
\label{cosmo.11}
\eeq
    The critical density is
\beq
    \rho_{c}=1.8\times
    10^{-23}h^{2}\frac{g}{cm^{3}}=10h^{2}\frac{nucleons}{m^{3}}
\label{cosmo.13}
\eeq
    With these definitions Friedman's equation
    $E=-\kappa\frac{m}{2}$
    takes the form
\beq
    \frac{\kappa}{a^{2}}=(\Omega _{m} +\Omega _{\Lambda}-1)H^{2}
\label{cosmo.14}
\eeq
    Thus we distinguish the following special cases:
\beq
    \kappa >0~~~~\Leftrightarrow~~~~\Omega _{m} +\Omega
    _{\Lambda}>1~~~~\Leftrightarrow Closed~curved~Universe
\eeq
\beq
    \kappa =0~~~~\Leftrightarrow~~~~\Omega _{m} +\Omega
    _{\Lambda}=1~~~~\Leftrightarrow Open~ Flat~ Universe
\eeq
\beq
    \kappa <0~~~~\Leftrightarrow~~~~\Omega _{m} +\Omega
    _{\Lambda}<1~~~~\Leftrightarrow Open~ Curved~ Universe
\eeq
   In other words it is the combination of matter and "vacuum"
   energy, which determines the fate of the our Universe.
                                                                                
\begin{figure}
\begin{center}
\includegraphics[height=0.8\textheight]{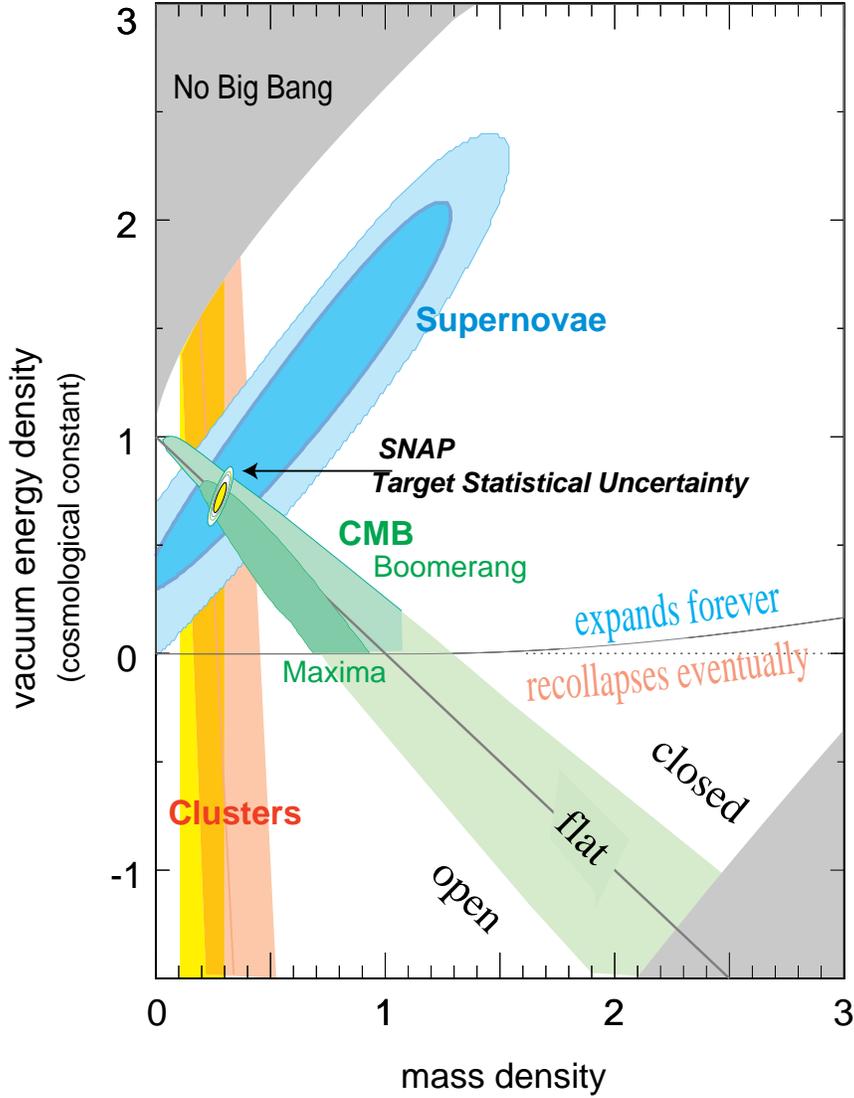}
\caption{The allowed region in the $\Omega_m$, $\Omega_{DE}$, with $ m$ indicating all matter and
$ DE$ the dark energy, for the various experiments indicated on the plot.
\label{OmLam} }
\end{center}
\end{figure}

    To proceed further and obtain $a(t)$ we need an equation of state, which relates the
pressure and the density. This of the form:
     $$p=w\rho c^2$$
the one finds
\beq
a(t)=const~t^{2/(3(1+w))}~~,~~\frac{\ddot{a}(t)}{a(t)}=-\frac{4\pi G_N}{3}(1+3w)\rho
\label{a(t)}
\eeq
We thus see that we have acceleration if $w<-1/3$ and deceleration otherwise.
 The following cases are of interest:
\begin{itemize}
\item $w=1/3$. This corresponds to a universe dominated by radiation
\item $w=0$. This corresponds to a universe dominated by non relativistic matter.
\item $w=-1$. This means that the cosmological constant is dominant. In this case the Hubble
 parameter is a constant and $a(t)=a_0 e^{H_0t}$. 
\item $-1<w<-1/3$. This corresponds to dark energy.
\end{itemize}
 We are not going to elaborate further on this issue. The observational situation at
present is summarized in Fig. \ref{OmLam}.

\section{Rates}
The differential non directional  rate can be written as
\begin{equation}
dR_{undir} = \frac{\rho (0)}{m_{\chi}} \frac{m}{A m_N}
 d\sigma (u,\upsilon) | {\boldmath \upsilon}|
\label{2.18}
\end{equation}
where $d\sigma(u,\upsilon )$ was given above,
$\rho (0) = 0.3 GeV/cm^3$ is the LSP density in our vicinity,
 m is the detector mass and
 $m_{\chi}$ is the LSP mass

 The directional differential rate, in the direction $\hat{e}$ of the
 recoiling nucleus, is:
\beq
dR_{dir} = \frac{\rho (0)}{m_{\chi}} \frac{m}{A m_N}
|\upsilon| \hat{\upsilon}.\hat{e} ~\Theta(\hat{\upsilon}.\hat{e})
 ~\frac{1}{2 \pi}~
d\sigma (u,\upsilon\
\nonumber \delta(\frac{\sqrt{u}}{\mu_r \upsilon
\sqrt{2}}-\hat{\upsilon}.\hat{e})
 \label{2.20}
\eeq

%
where $\Theta(x)$ is the Heaviside function and: 
\beq                                                                                         
d\sigma (u,\upsilon)== \frac{du}{2 (\mu _r b\upsilon )^2}
 [(\bar{\Sigma} _{S}F(u)^2
                       +\bar{\Sigma} _{spin} F_{11}(u)]
\label{2.9}
\end{equation}
where $ u$ the energy transfer $Q$ in dimensionless units given by
\begin{equation}
 u=\frac{Q}{Q_0}~~,~~Q_{0}=[m_pAb]^{-2}=40A^{-4/3}~MeV
\label{defineu}
\end{equation}
 with  $b$ is the nuclear (harmonic oscillator) size parameter. $F(u)$ is the
nuclear form factor and $F_{11}(u)$ is the spin response function associated with
the isovector channel. It is the area of constructing such form factors and momentum
distributions that Professor M. Grypeos, honored in this volume,
 has made a great contribution \cite{grypeos}.
 
The scalar
cross section is given by:
\begin{equation}
\bar{\Sigma} _S  =  (\frac{\mu_r}{\mu_r(p)})^2
                           \sigma^{S}_{p,\chi^0} A^2
 \left [\frac{1+\frac{f^1_S}{f^0_S}\frac{2Z-A}{A}}{1+\frac{f^1_S}{f^0_S}}\right]^2
\approx  \sigma^{S}_{N,\chi^0} (\frac{\mu_r}{\mu_r (p)})^2 A^2
\label{2.10}
\end{equation}
(since the heavy quarks dominate the isovector contribution is
negligible). $\sigma^S_{N,\chi^0}$ is the LSP-nucleon scalar cross section.
The spin Cross section is given by:
\begin{equation}
\bar{\Sigma} _{spin}  =  (\frac{\mu_r}{\mu_r(p)})^2
                           \sigma^{spin}_{p,\chi^0}~\zeta_{spin},
\zeta_{spin}= \frac{1}{3(1+\frac{f^0_A}{f^1_A})^2}S(u)
\label{2.10a}
\end{equation}
\begin{equation}
S(u)\approx S(0)=[(\frac{f^0_A}{f^1_A} \Omega_0(0))^2
  +  2\frac{f^0_A}{ f^1_A} \Omega_0(0) \Omega_1(0)+  \Omega_1(0))^2  \, ]
\label{s(u)}
 \end{equation}
Some static of interest to the planned experiments are given in table \ref{table.spin}.
 The shown results
are obtained from DIVARI \cite{DIVA00}, Ressel and Dean (*) \cite{Ress},
 the Finish group (**) \cite {SUHONEN03} and the Ioannina team (+) \cite{ref1}, \cite{KVprd}.
 
\begin{table}[t]
\caption{
 The static spin matrix elements for various nuclei. For
light nuclei the calculations are from DIVARI.
 For $^{127}I$ the results presented  are from Ressel and Dean
(*) and the Finish group 
 (**).
 For $^{207}Pb$ they were obtained by the Ioannina team (+). 
\label{table.spin}
}
\begin{center}
\begin{tabular}{lrrrrrr}
\hline\hline
 &   &  &  &  &   & \\
 & $^{19}$F & $^{29}$Si & $^{23}$Na  & $^{127}I^*$ & $ ^{127}I^{**}$ & $^{207}Pb^+$\\
\hline
    &   &  &  &  &    \\
$[\Omega_{0}(0)]^2$         & 2.610   & 0.207  & 0.477  & 3.293   &1.488 & 0.305\\
$[\Omega_{1}(0)]^2$         & 2.807   & 0.219  & 0.346  & 1.220   &1.513 & 0.231\\
$\Omega_{0}(0)\Omega_{1}(0)$& 2.707   &-0.213  & 0.406  &2.008    &1.501&-0.266\\
$\mu_{th} $& 2.91   &-0.50  & 2.22  &    &\\
$\mu_{exp}$& 2.62   &-0.56  & 2.22  &    &\\
$\frac{\mu_{th}(spin)}{ \mu_{exp}}$& 0.91   &0.99  & 0.57  &    &  &\\
\hline
\hline
\end{tabular}
\end{center}
\end{table}

 $ f^0_A$, $f^1_A$ are the isoscalar and the isovector axial current
couplings at the nucleon level obtained from the corresponding ones given by the SUSY
 models at the quark level, $ f^0_A(q)$, $f^1_A(q)$, via renormalization
coefficients $g^0_A$, $g_A^1$, i.e.
$ f^0_A=g_A^0 f^0_A(q),f^1_A=g_A^1 f^1_A(q).$
These couplings and the associated nuclear matrix elements are normalized so that,
 for the proton at $u=0$, yield $\zeta_{spin}=1$. If the nuclear contribution
 comes predominantly from protons
 ($\Omega_1=\Omega_0=\Omega_p$), $S(u)\approx\Omega_p^2$ and one can extract from
the data the proton cross section. If the nuclear contribution comes predominantly
from neutrons (($\Omega_0=-\Omega_1=\Omega_n$) one can extract the neutron cross
section.
 In many cases, however, one can have contributions from both protons and
neutrons. The situation is then complicated, but it
 turns out that $g^0_A=0.1~~,~~g^1_A=1.2$
Thus the isoscalar amplitude
 is suppressed, i.e  $S(0)\approx \Omega^2_1$.
Then the proton and the neutron  spin cross sections are the same.

\section{Results}
To obtain the total rates one must fold with LSP velocity and integrate  the 
above expressions  over the
energy transfer from $Q_{min}$ determined by the detector energy cutoff to $Q_{max}$
determined by the maximum LSP velocity (escape velocity, put in by hand in the
Maxwellian distribution), i.e. $\upsilon_{esc}=2.84~\upsilon_0$ with  $\upsilon_0$
the velocity of the sun around the center of the galaxy($229~Km/s$).

\subsection{Non directional rates}
Ignoring the motion of the Earth the 
total non directional rate is given by
\begin{equation}
R= \bar{K} \left[c_{coh}(A,\mu_r(A)) \sigma_{p,\chi^0}^{S}+
c_{spin}(A,\mu_r(A))\sigma _{p,\chi^0}^{spin}~\zeta_{spin} \right]
\label{snew}
\end{equation}
where $\bar{K}=\frac{\rho (0)}{m_{\chi^0}} \frac{m}{m_p}~
              \sqrt{\langle v^2 \rangle } $ and
\begin{equation}
c_{coh}(A, \mu_r(A))=\left[ \frac{\mu_r(A)}{\mu_r(p)} \right]^2 A~t_{coh}(A)~,
c_{spin}(A, \mu_r(A))=\left[ \frac{\mu_r(A)}{\mu_r(p)} \right]^2 \frac{t_{spin}(A)}{A}
\label{ctm}
\end{equation}
 where $t$ is the modification of the total rate due to the
 folding and nuclear structure effects. It depends on
$Q_{min}$, i.e.  the  energy transfer cutoff imposed by the detector
 and $a=[\mu_r b \upsilon _0 \sqrt 2 ]^{-1}$
The parameters $c_{coh}(A,\mu_r(A))$, $c_{spin}(A,\mu_r(A))$, which give the relative merit 
 for the coherent and the spin contributions in the case of a nuclear
target compared to those of the proton,  are tabulated in table \ref{table.murt}
 for 
energy cutoff $Q_{min}=0,~10$ keV.
Via  Eq. (\ref{snew}) we can  extract the nucleon cross section from
 the data.
 
\begin{figure}
\begin{center}
\includegraphics[height=0.4\textheight]{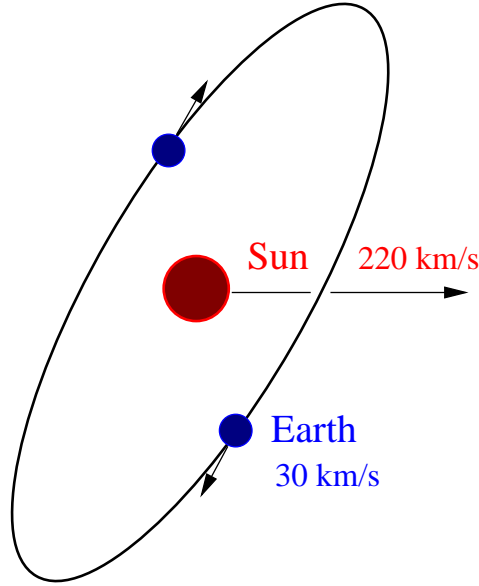}
\caption{The motion of the earth around the sun. The sun is moving with a velocity in the plane of
the galaxy, which is perpendicular to the page.
\label{solar} }
\end{center}
\end{figure}

\begin{figure}
\begin{center}
\rotatebox{90}{\hspace{1.0cm} $h\rightarrow$}
\includegraphics[height=0.16\textheight]{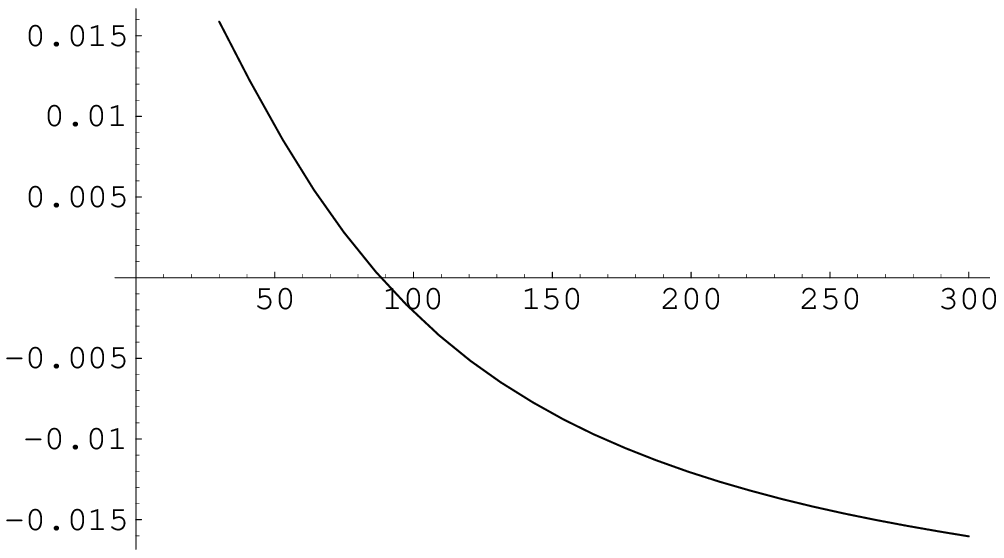}
\rotatebox{90}{\hspace{1.0cm} $h\rightarrow$}
\includegraphics[height=0.16\textheight]{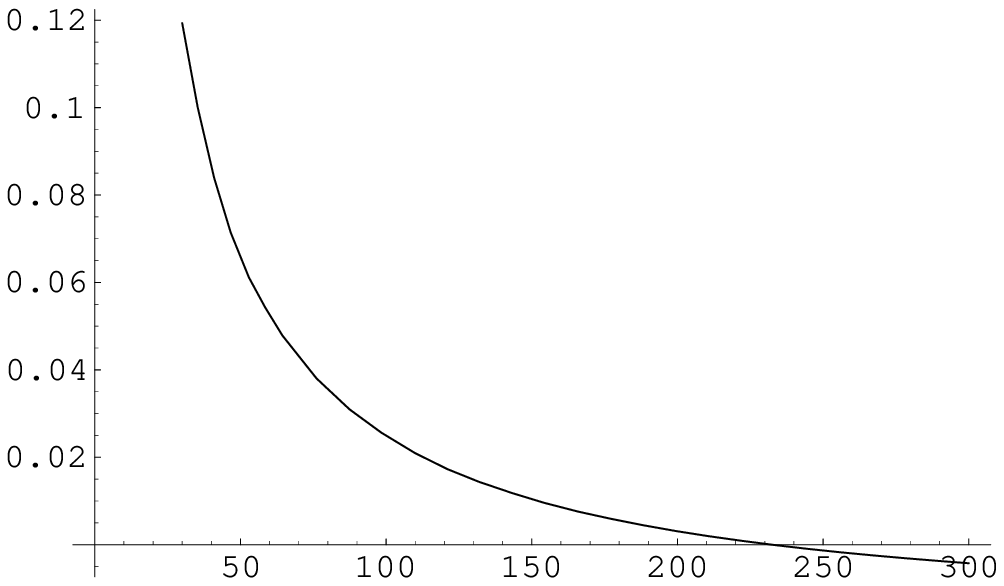}

\hspace{2.0cm} $m_{LSP}\rightarrow$ (GeV)
\caption{The modulation
amplitude $h$ as a function of the LSP mass in the case of $^{127}$I for
$Q_{min}=0$ on the left and $Q_{min}=10$ keV on the right. 
\label{hgs} }
\end{center}
\end{figure}

 Using
$\Omega^2_1=1.22$ and $\Omega^2_1=2.8$ for $^{127}$I and $^{19}$F respectively
the extracted nucleon cross sections satisfy: 
\begin{equation}
\frac{\sigma^{spin}_{p,\chi^0}}{\sigma^{S}_{p,\chi^0}} =
 \left[\frac{c_{coh}(A,\mu_r(A))}{c_{spin}(A,\mu_r(A))}\right]
\frac{3 }{\Omega^2_1} \Rightarrow
\approx  \times 10^{4}~(A=127)~,~ \approx \times 10^2~(A=19)
\label{ratior2}
\end{equation}
It is for this reason that the limit on the spin proton cross section extracted from both
targets is much poorer.
 For heavy LSP, $\ge 100$ GeV, due to the nuclear form factor,
 $t_{spin}(127)< t_{spin}(19)$.  This disadvantage 
 cannot be overcome by the larger reduced mass. It even becomes worse,
  if the effect
of the spin ME is included. For the coherent process, however,
the light nucleus is no match
 (see Table \ref{table.murt}). 
\begin{table}[t]
\caption{
The factors $c19= c_{coh}(19,\mu_r(19))$,  $s19= c_{spin}(19,\mu_r(19))$
 and 
$c127=c_{coh}(127,\mu_r(127))$,  $s127= c_{spin}(127,\mu_r(127))$
for two values of $Q_{min}$.
\label{table.murt}
}
\begin{center}
\begin{tabular}{|r|r|rrrrrrrr|}
\hline
$Q_{min}$& &\multicolumn{8}{c|}{$m_{\chi}$ (GeV)}\\
\hline
& &   &  &  &  &   & & &\\
keV& & 20 & 30 & 40  & 50 & 60 & 80&100&200\\
\hline
0&c19&2080&2943&3589&4083&4471&5037&5428&6360\\
0&s19&5.7&8.0&9.7&10.9&11.9&13.4&14.4&16.7\\
\hline
0&c127&37294&63142&84764&101539&114295&131580&142290&162945\\
0&s127&2.2&3.7&4.9&5.8&6.5&7.6&8.4&10.4\\
\hline
10&c19&636&1314&1865&2302&2639&3181&3487&4419\\
10&s19&1.7&3.5&4.9&6.0&6.9&8.3&9.1&11.4\\
\hline
10&c127&0&11660&24080&36243&45648&58534&69545&83823\\
10&s127&0&0.6&1.3&1.9&2.5&3.3&4.0&5.8\\
\hline
\end{tabular}
\end{center} 
\end{table}
\\If the effects of the motion of the Earth around the sun are included, the total
 non directional rate is given by
\begin{equation}
R= \bar{K} \left[c_{coh}(A,\mu_r(A)) \sigma_{p,\chi^0}^{S}(1 +  h(a,Q_{min})cos{\alpha})\right] 
\label{3.55j} 
\end{equation} 
and an analogous one for the spin contribution.  $h$  is the modulation amplitude and
 $\alpha$ is the phase of the Earth, which is
zero around June 2nd, see Fig \ref{solar}.
 The modulation amplitude would be an excellent signal in
discriminating against background, but unfortunately it is very small, less than two 
per cent. Furthermore for intermediate and heavy nuclei, it can even change sign
for sufficiently  heavy LSP (see Fig. \ref{hgs}).
 So in our opinion a better signature is provided
 by directional experiments, which measure the direction of the recoiling nucleus.

\subsection{Directional Rates.}
\begin{figure}[htb]
\begin{center}
\includegraphics[height=.15\textheight]{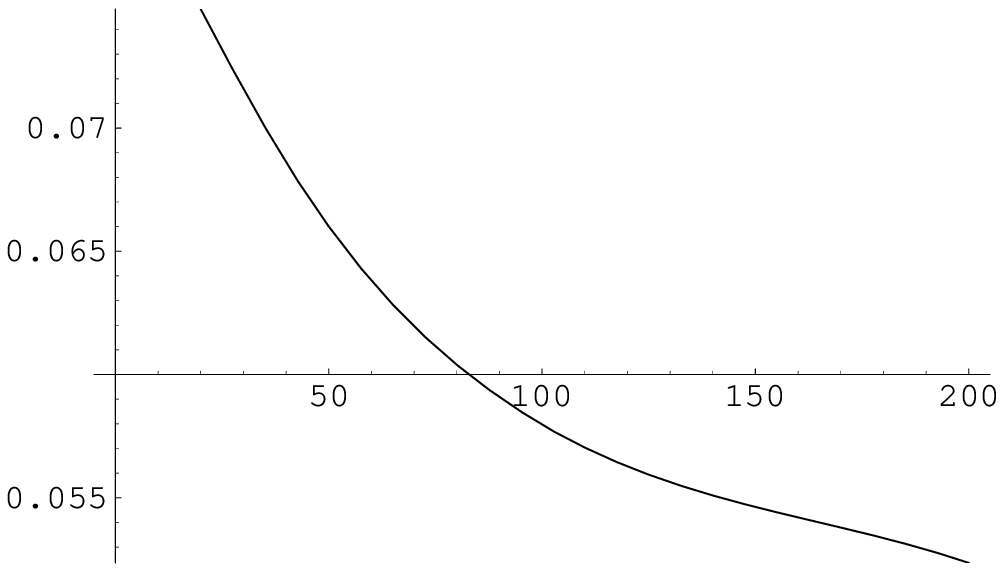}
\includegraphics[height=.15\textheight]{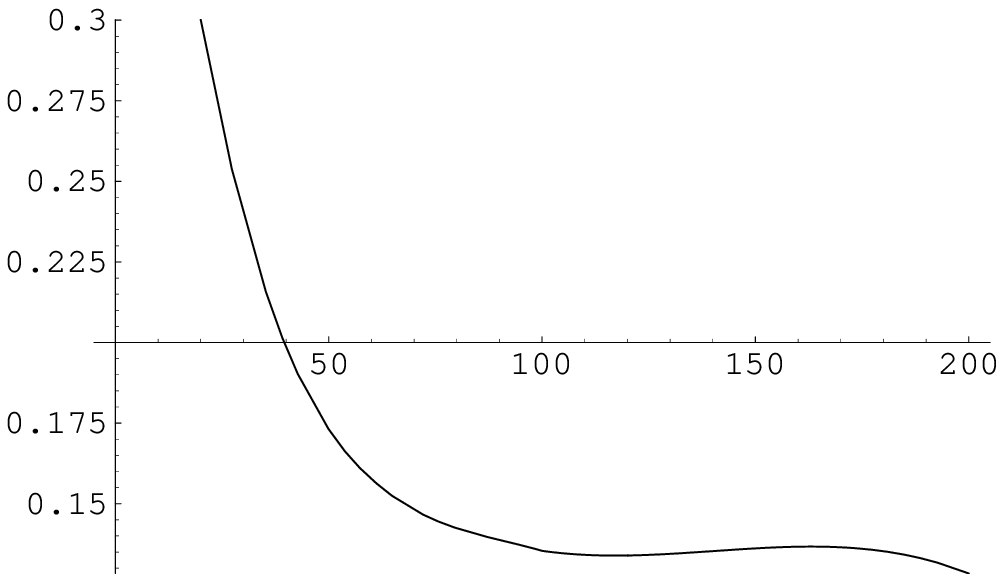}
\caption{ The parameter $\kappa$ as a function of the LSP mass in the case of
the $A=127$ system,
for $Q_{min}=0$ expected in a plane perpendicular to the sun's velocity on the left and opposite 
to the sun's velocity on the right.
 \label{k.127}
}
\end{center}
\end{figure}

\begin{figure}[htb]
\begin{center}
\includegraphics[height=.18\textheight]{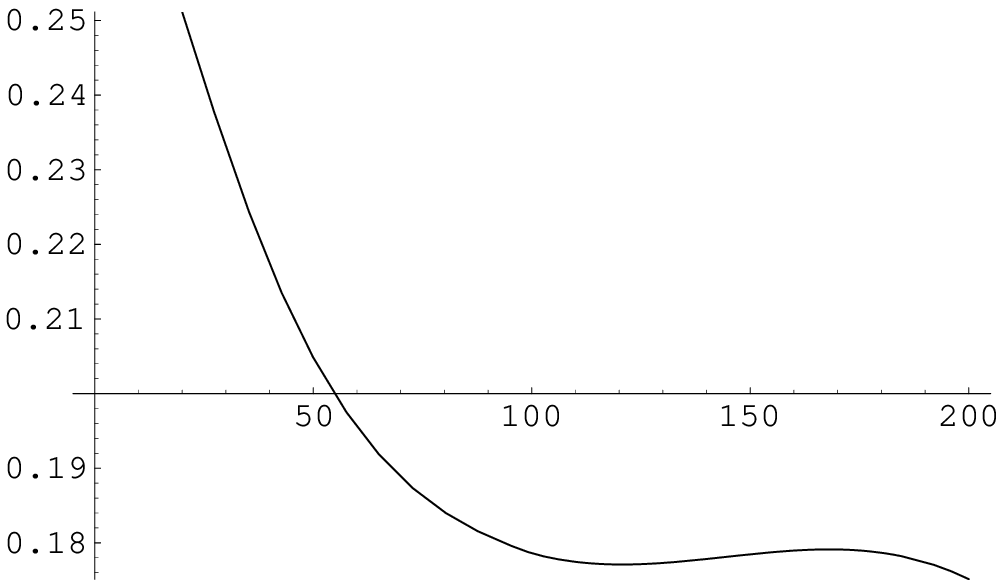}
\includegraphics[height=.18\textheight]{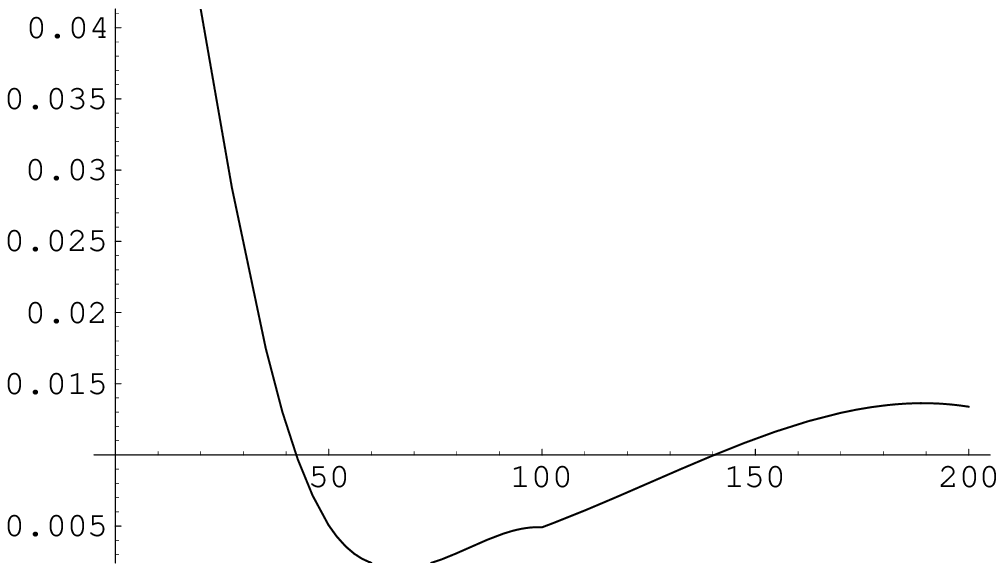}
\caption{ The modulation amplitude $h_m$
 in a plane perpendicular to the sun's velocity on the left and opposite to the suns velocity
on the right.
 Otherwise the notation is the same as in Fig \ref{k.127}.
 \label{h.127}
 }
\end{center}
\end{figure}

Since the sun is moving around the galaxy in a directional experiment, i.e. one in which the
direction of the recoiling nucleus is observed, one expects a strong correlation of the
event rate with the motion of the sun. In fact
the directional rate can be written as:
\begin{equation}
R_{dir}  = 
          \frac{\kappa} {2 \pi} \bar{K} \left[c_{coh}(A,\mu_r(A)) \sigma_{p,\chi^0}^{S} 
            (1 + h_m  cos {(\alpha-\alpha_m~\pi)}) \right ]
\label{4.56b}  
\end{equation}
and an analogous one for the spin contribution. The modulation now is  $h_m$, with 
a shift $\alpha_m \pi $ in the phase of the Earth $\alpha$, 
depending on the direction of observation.
 $\kappa/(2 \pi)$ is the reduction factor
of the unmodulated directional rate relative
to the non-directional one. 
The parameters  $\kappa~,~h_m~,~\alpha_m$ strongly depend on the direction of
 observation.
\begin{figure}[htb]
\begin{center}
\includegraphics[height=.15\textheight]{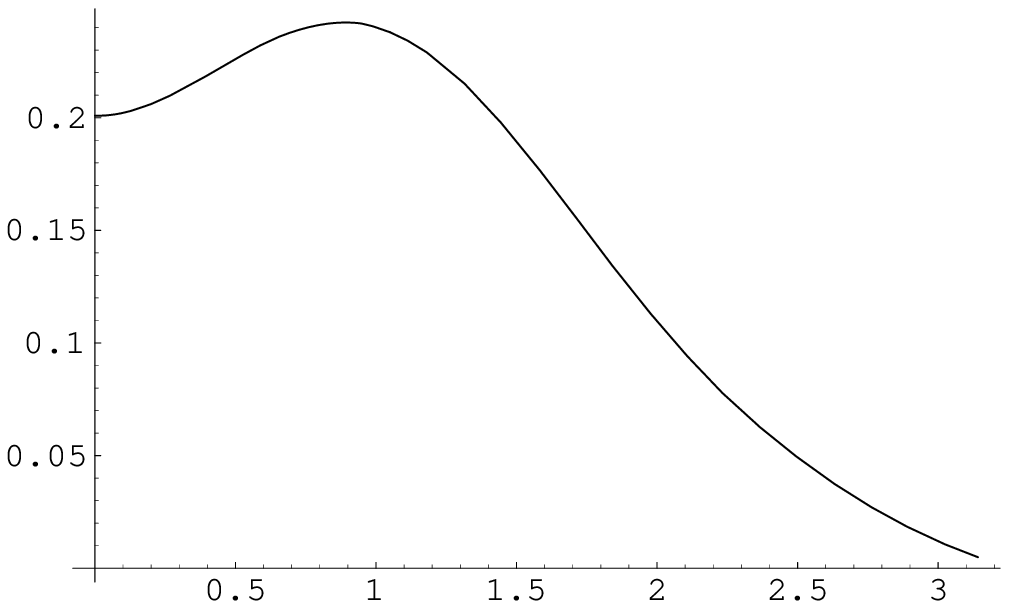}
\includegraphics[height=.15\textheight]{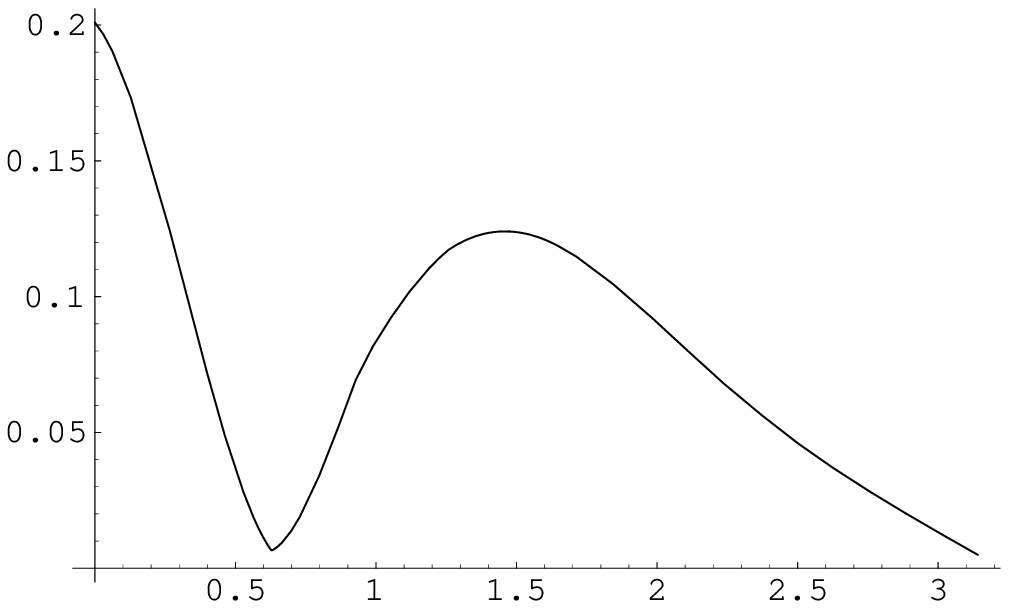}
\caption{ The expected modulation amplitude $h_m$ for $A=127$  in a direction outward
from the galaxy on the left and perpendicular to the galaxy on the right as a function 
of the polar angle measured from the sun's velocity. For angles than $\pi/2$ it is
 irrelevant
since the event rates are tiny.
 \label{hdir127}
 }
\end{center}
\end{figure}
We prefer to use the parameters $\kappa$ and $h_m$, since,
being ratios, are expected to be 
 less dependent on the parameters of the theory. In the case of $A=127$ we  exhibit the
the angular dependence of $h_m$ for an LSP mass of $m_{\chi}=100GeV$
in Fig.  \ref{hdir127}. We also exhibit  the
parameters $t$, $h$, $\kappa,h_m$ and $\alpha_m$ for the target $A=19$
in Table \ref{table1.gaus} (for the other light systems the results are
almost identical).
\begin{table}[t]  
\caption{ The parameters $t$, $h$, $\kappa,h_m$ and $\alpha_m$ for 
  $Q_{min}=0$.
 The results shown are for the light
systems. $+x$ is
 radially  out of the galaxy , $+z$ is in the  the sun's
 motion  and
$+y$ vertical to the plane of the galaxy  so that 
 $((x,y,x)$ is right-handed. $\alpha_m=0,1/2,1,3/2$ implies
that the maximum occurs on June, September, December and March 2nd
 respectively.
\label{table1.gaus}
}
\begin{center}
\begin{tabular}{|lrrrrrr|}
& & & & & &      \\
type&t&h&dir &$\kappa$ &$h_m$ &$\alpha_m$ \\
\hline
& & & & & &      \\
& &&+z        &0.0068& 0.227& 1\\
dir& & &+(-)x      &0.080& 0.272& 3/2(1/2)\\ 
& & &+(-)y        &0.080& 0.210& 0 (1)\\
& & &-z         &0.395& 0.060& 0\\
\hline
all&1.00& & && & \\
all& & 0.02& & & & \\
\hline
\end{tabular}
\end{center}
\end{table}
\\
The asymmetry in the direction of the sun's motion \cite{JDV03} is quite large,
$\approx 0.97$, while in the  perpendicular plane
the asymmetry equals the modulation.\\
For a heavier nucleus the situation is a bit complicated. Now the
parameters $\kappa$ and $h_m$ depend on the LSP mass as well. 
 (see Figs \ref{k.127} and \ref{h.127}). The asymmetry and the shift
in the phase of the Earth are similar to those of the $A=19$ system.

\subsection{Transitions to excited states}
Incorporating the relevant kinematics  and integrating
  the differential event rate $dR/du$
 from $u_{min}$ to $u_{max}$ we
 obtain the total rate as follows:
 \beq
 R_{exc}=\int_{u_{exc}}^{u_{max}}\frac{dR_{exc}}{du}(1-\frac{u^2_{exc}}{u^2})du~,
~ R_{gs}=\int_{u_{min}}^{u_{max}}\frac{dR_{gs}}{du}du
 \label{Rexc}
 \eeq
 where $u_{exc}=\frac{\mu_rE_{x}}{Am_NQ_0}$ and $E_{x}$ is the
 excitation energy of the final nucleus, $u_{max}=(y/a)^2-(E_x/Q_0)$
 , $y=\upsilon/upsilon_0$ and
 $u_{min}=Q_{min}/Q_0$, $Q_{min}$ (imposed by the detector
 energy cutoff) and $u_{max}=(y_{esc}/a)^2$
 is imposed by the escape velocity ($y_{esc}=2.84$).

\begin{figure}
\begin{center}
\rotatebox{90}{\hspace{1.0cm} {\tiny BRR}$\rightarrow$}
\includegraphics[height=.18\textheight]{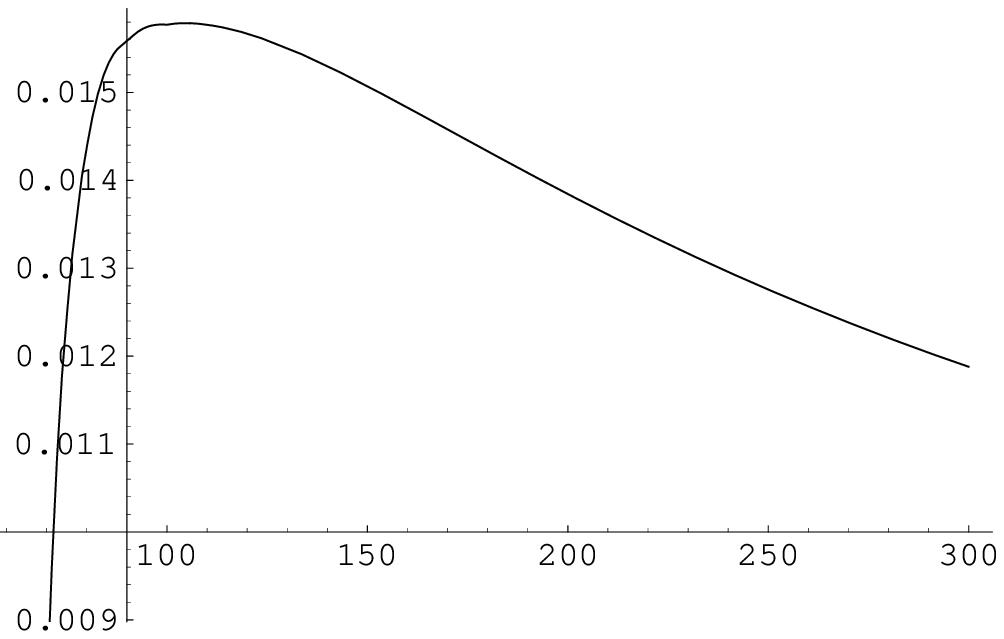}
 \rotatebox{90}{\hspace{1.0cm} {\tiny BRR}$\rightarrow$}
\includegraphics[height=.18\textheight]{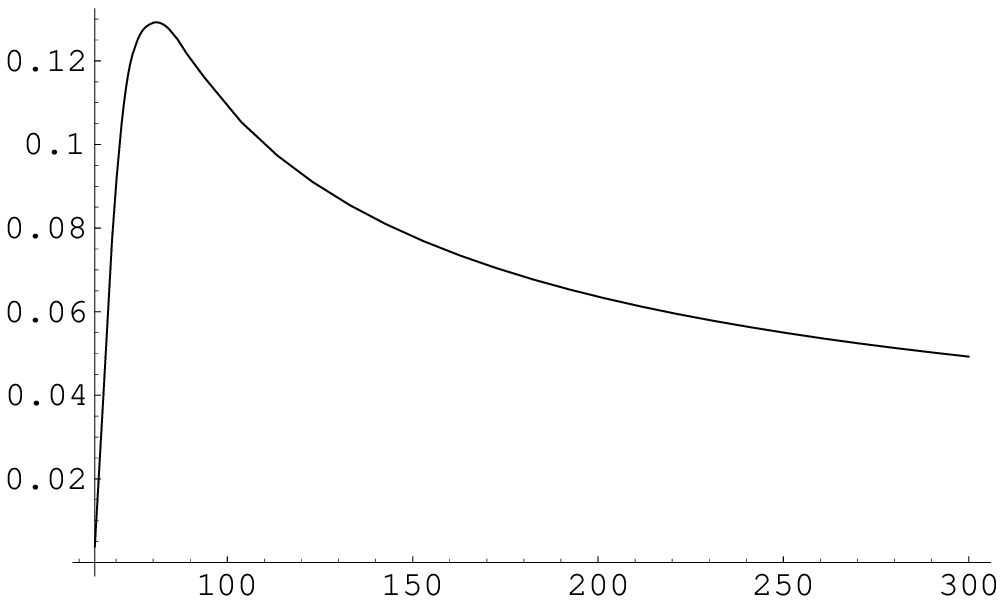}

 \hspace{0.0cm} $m_{LSP}\rightarrow$ ($GeV$)
 \caption{ The ratio of the rate to
the excited state divided by that of the ground state as a
function of the LSP mass (in GeV) for $^{127}I$. We assumed that
the static spin matrix element of the transition from the ground
to the excited state is a factor of 1.9 larger than that
involving the ground state, but  the  functions $F_{11}(u)$ are
the same. On the left we show the results for $Q_{min}=0$ and on
the right for $Q_{min}=10~KeV$. \label{ratio} }
\end{center}
\end{figure}

For our purposes
it is adequate to estimate the ratio of the rate
 to the excited state divided by that to the ground state
 (branching ratio) as a function of the LSP mass.
  This can be cast in the form:
\begin{equation}
BRR =  \frac{S_{exc}(0)}{S_{gs}(0)}
    \frac{\Psi_{exc}(u_{exc},u_{umax})[1+h_{exc}(u_{exc},u_{max})~\cos{\alpha}]}
           {\Psi_{gs}(u_{min})[1+h(u_{min})~\cos{\alpha}]}
\label{rR}
\end{equation}
in an obvious notation \cite{JDV03}).
 $S_{gs}(0)$ and $S_{exc}(0)$ are the static spin matrix elements
As we have seen their ratio
is essentially independent of supersymmetry, if the isoscalar
contribution is neglected. For $^{127}$I it was found \cite{VQS03}
to be be about 2. The functions
$\Psi$ are given as follows :
\begin{equation}
\Psi_{gs} (u_{min}) =\int _{u_{min}}^{(y/a)^2} \frac{S_{gs}
(u)}{S_{gs} (0)} F^{gs}_{11}(u)
             \big[ \psi (a \sqrt{u}) - \psi (y_{esc}) \big] ~du
\label{Psigs}
\end{equation}
\barr
\Psi_{exc} (u_{exc},u_{max})& =&\int _{u_{exc}}^{u_{max}}
\frac{S_{exc} (u)}{S_{exc} (0)}
F^{exc}_{11}(u)(1-\frac{u^2_{exc}}{u^2})
\nonumber\\
& & \big[ \psi (a \sqrt{u}(1+u_{exc}/u)) - \psi (y_{esc}) \big] ~du
\label{Psi1}
\earr
The functions $\psi$ arise from the convolution with LSP velocity distribution.
The obtained results are shown in Fig. \ref{ratio}.
\begin{figure}
\begin{center}
\rotatebox{90}{\hspace{1.0cm} $h_{exc}\rightarrow$}
\includegraphics[height=.18\textheight]{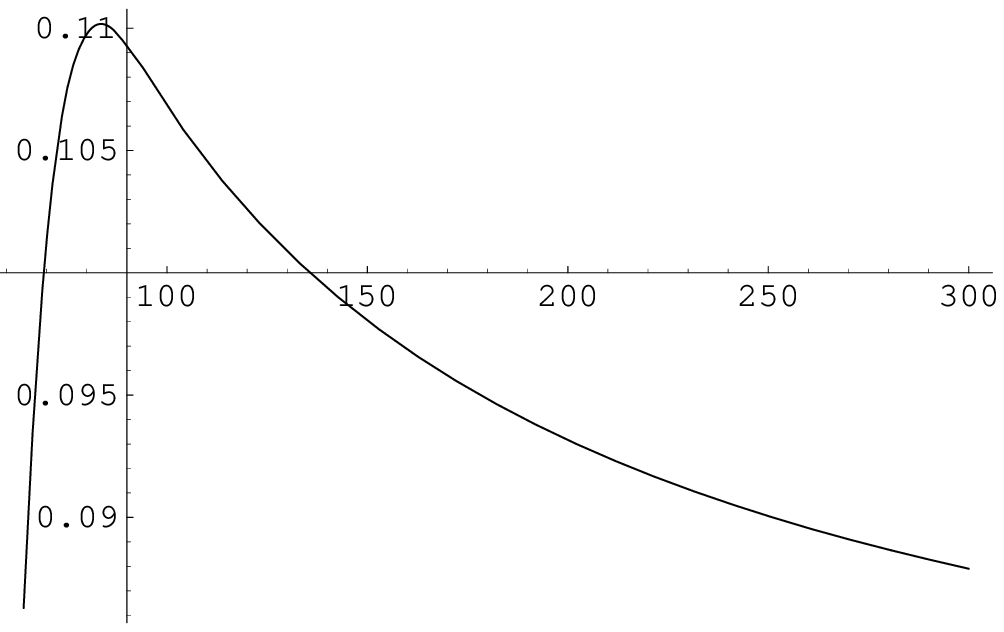}
\rotatebox{90}{\hspace{1.0cm} $h_{exc}\rightarrow$}
\includegraphics[height=.18\textheight]{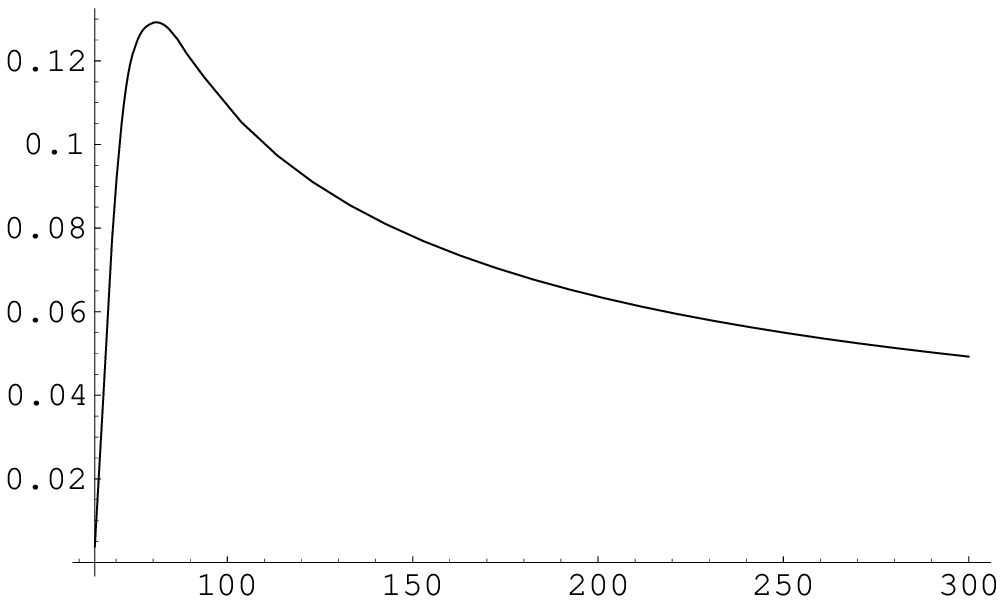}

\hspace{0.0cm} $m_{LSP}\rightarrow$ ($GeV$)
 \caption{ The the same
as in Fig. \ref{hgs} for the modulation amplitude $h_{exc}$ for
the transition to the excited state. \label{hexc} }
\end{center}
\end{figure}

\subsection{Detecting recoiling electrons following the LSP-nucleus collision.}
 During the LSP-nucleus collision we can have the ionization of the atom. One thus
may exploit this signature of the reaction and try to detect these electrons
\cite{VEJ04}. These electrons
are expected to be of low energy and  one thus may have a better chance of
observing them in a gaseous TPC detector. In order to avoid complications arising from
atomic physics we have chosen as a target $^{20}$Ne. Furthermore, to  avoid  complications regarding the
 allowed SUSY parameter space,
   we will present our results normalized to
  the standard neutralino nucleus cross section.
   The thus obtained branching  ratios are
   independent of all parameters of supersymmetry except the neutralino
   mass. The numerical results given here apply in the case of the coherent
   mode. If, however, we limit ourselves to the ratios of the
   relevant cross sections,
    we do not expect  substantial changes in the case of the
  spin induced process.
                                                                                                   
  The ratio of our differential (with respect of the electron energy) cross
 section divided by the total cross section of the standard
 neutralino-nucleus elastic scattering, nuclear recoil experiments (nrec), takes
\cite{VEJ04} the form:
 \barr
 \frac {d\sigma(T)}{\sigma_{nrec}}&=&\frac{1}{4}\sum_{n\ell}p_{n\ell}
|\tilde{\phi_{n\ell}}(2m_eT)|^2
  \nonumber\\
  && \frac{\int _{-1} ^1
d\xi_1\int_{\xi_L} ^1 d \xi K \frac{(\xi+\Lambda)^2}{\Lambda}
 [F(\mu_r\upsilon (\xi+\Lambda))]^2}{\int_0^1 2 \xi d \xi
 [F(2 \mu_r \upsilon \xi)]^2}{m_e}k  dT.
 \label{ratio1}
  \earr
where $${\bf K}=\frac{{\bf p}_{\chi}-{\bf k}}{p_{\chi}},
K=\frac{\sqrt{p_{\chi}^2+k^2-2k p_{\chi} \xi_1}}{p_{\chi}},
\xi_1=\hat{p}_{\chi}.\hat{k},\xi=\hat{q}.\hat{K},$$
 $\xi_L=
\sqrt{\frac{m_{\chi}}{\mu_r}[1+\frac{1}{K^2}(\frac{T-\epsilon_{n\ell}}{T_{\chi}}-1)]}$
and
$2\frac{\mu_r}{m_{\chi}} p_{\chi}\xi=2\mu_r\upsilon\xi$ is
 the momentum $q$ transferred to the nucleus and
             the momentum $q$ transferred to the nucleus and
 $F(q)$ is the nuclear form factor. The outgoing electron energy
lies in the range $0\leq T\leq
\frac{\mu_r}{m_{\chi}}T_{\chi}-\epsilon_{n\ell}$.
Since the momentum of the outgoing electron is much smaller than
the momentum of the oncoming neutralino, i.e. $K\approx 1$, the
integration over $\xi_1$ can be trivially performed.
                                                                                                   
\begin{figure}
\rotatebox{90}{\hspace{0.0cm}
 $\frac{1}{R}\frac{dR_e}{ dT}\rightarrow keV^{-1}$}
\includegraphics[height=.13\textheight]{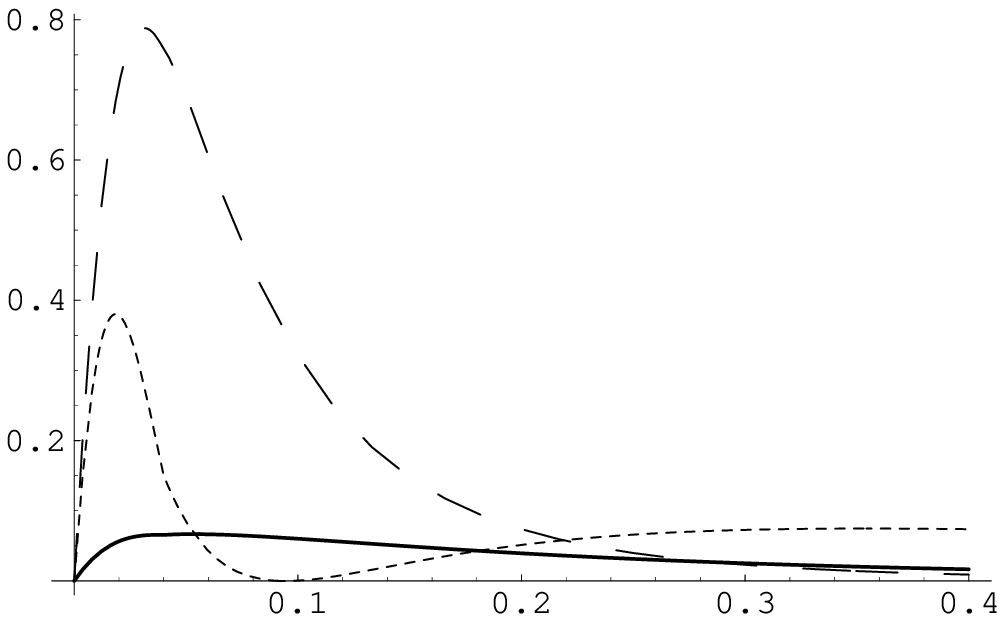}
{\hspace{0.0cm} {\tiny $T\rightarrow keV $}}
\rotatebox{90}{\hspace{0.0cm}
 $\frac{R_e}{R}\rightarrow$}
\includegraphics[height=.12\textheight]{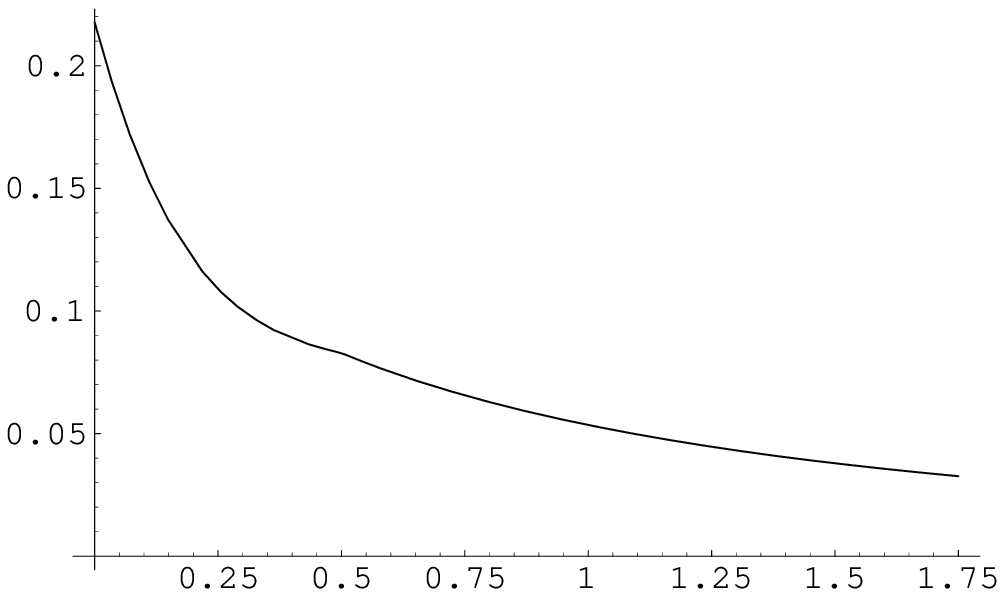}
{\hspace{0.0cm} {\tiny $E_{th}\rightarrow keV $}}
\caption{Shown on the left is the differential rate, divided by the total rate
associated with the nuclear recoils, as a function of the electron
energy T (in $keV$). Each atomic orbit involved in the target
$^{20}$Ne is included separately.
 The full line, the short-dashed line and the
long-dashed line correspond to the orbits $1s~,~2s~$ and $2p$
respectively. Shown on the right is the ratio of the total rate for the novel
process divided by that of the standard process as a function of the electron
threshold energy, assuming zero threshold energy for the standard process. This ratio
may increase if such a threshold is included.
 \label{rationew}}
\end{figure}

  We remind the reader that the LSP- nucleus cross-section $\sigma_{nrec}$
takes the form:
  \beq
\sigma_{nrec}=(\frac{\mu_r}{\mu_r (p)})^2 A^2 \sigma_p
 \int_0^1 2 d \xi [F(2\mu_r \upsilon \xi)]^2
 \label{slsp}
  \eeq
 In the case of $^{20}$Ne he binding energies and the occupation probabilities
are given by\cite{VEJ04}:
\beq
\epsilon_{n\ell}=(-0.870,-0.048~,-0.021)~,~   p_{n\ell}=(2/10,2/10,6/10).
 \label{bener}
 \eeq
in the obvious order: (1s,2s,2p).
In Fig.\ref{rationew} we show the differential rate of our process,
 divided by the total  nuclear recoil event rate,
 for each orbit
  as well as the total rate in our process divided
by that of the standard rate as a function of the electron threshold energy
with $0$ threshold energy in the standard process. We obtained our results
using  appropriate form
 factor \cite{DIVA00}.
                                                                                                   
 From these plots we see that, even though the differential rate peaks at low energies,
there remains
 substantial strength above the electron energy of $0.2~keV$, which is the threshold
 energy for
 detecting electrons in a Micromegas detector, like the one recently
 \cite{GV03} proposed.
\section{Conclusions}
Since the expected event rates for direct neutralino detection are
very low\cite{ref2,ARNDU},
 in the present work we  looked for
characteristic experimental signatures for background reduction, such as:
\\{\bf Standard recoil experiments.}
 Here  the relevant parameters  are $t$ and $h$. For light targets
they  are essentially independent of the LSP mass \cite{JDV03},
  essentially the same for both the coherent and the spin modes.
The modulation is small, $h \approx 0.2\%$, but it  may increase as
$Q_{min}$ increases. Unfortunately, for heavy targets even the sign of $h$
 is uncertain for $Q_{min}=0$. The
situation improves  as $Q_{min}$ increases, but at the expense of the number of counts.
\\{\bf Directional experiments} \cite{DRIFT}.
Here we find a correlation of the rates with the velocity of the sun
 as well as that of the Earth.
One encounters reduction factors
 $\kappa/2\pi$, which depend on the angle of observation. The most favorable factor 
is small, $\approx1/4\pi$ and occurs when the nucleus is recoiling
opposite to the direction of motion of the sun. As a bonus  one gets  modulation, which is
three times larger, $h_m=\approx 0.06$.
 In a plane perpendicular to the sun's direction of motion the reduction
factor is close to $1/12\pi$, but now the modulation can be quite high, $h_m\approx0.3$,
 and exhibits
very interesting time dependent pattern (see Table \ref{table1.gaus}.
 Further interesting features may appear in the case of non standard velocity 
distributions \cite{GREEN04}.
\\ {\bf Transitions to Excited states.}
 We find that branching ratios for transitions to the first excited state of
$^{127}$I is relatively high, about $10\%$. The modulation in this case
is much larger $h_{exc}\approx 0.6$.
We hope that such a branching ratio will encourage future 
experiments to search for characteristic $\gamma$ rays rather than recoils.
\\{\bf Detection of ionization electrons produced in the LSP collision}                              
Our  results indicate that one can be optimistic
   about using the emitted electrons in the neutralino nucleus
   collisions for the direct detection of the LSP.  This novel process
   may be exploited by the planned TPC low energy electron
   detectors. By achieving   low energy thresholds of about $0.25~keV$, the
   branching ratios are approximately 10 percent. They can be even
   larger,
   if one includes low energy cutoffs imposed by the detectors in the
   standard experiments, not included in the above estimate.
                                                                                                   
    As we have seen the background problems associated with the proposed
    mechanism are not worse than those entering the standard experiments. In
    any case coincidence experiments with
   x-rays, produced following the de-excitation of the residual atom, may help reduce
   the background events to extremely low levels.

Acknowledgments: This work was supported in part by the
European Union under the contracts RTN No HPRN-CT-2000-00148 and
MRTN-CT-2004-503369. Part of this work was performed in LANL.
The author is indebted to Dr Dan Strottman
for his support and hospitality.


\begin{thebibliography}{99}
\bibitem{ALLEXP}
S. Hanary {\it et al}, {\it Astrophys. J.} {\bf 545}, L5 (2000);
J.H.P Wu {\it et al}, {\it Phys. Rev. Lett.} {\bf 87}, 251303 (2001);
M.G. Santos {\it et al}, {\it ibid} {\bf 88}, 241302
(2002)\\
P.D. Mauskopf {\it et al}, {\it Astrophys. J.} {\bf 536}, L59 (20002);
S. Mosi {\it et al}, {\it Prog. Nuc.Part. Phys.} {\bf 48}, 243 (2002);
S.B. Ruhl {\it al}, astro-ph/0212229;
N.W. Halverson {\it et al}, {Astrophys. J.} {\bf 568}, 38 (2002);
L.S. Sievers {\it et al},astro-ph/0205287;
G.F. Smoot {\it et al}, (COBE data), {\it Astrophys. J.} {\bf
396}, (1992) L1;
A.H. Jaffe {\it et al.},{\it Phys. Rev. Lett.} {\bf 86}, 3475
(2001);
D.N. Spergel et al, astro-ph/0302209
\bibitem{Jungm}
G. Jungman, M. Kamiokonkowski and K. Griest, Phys. Rep. {\bf 267}, 195 (1996).
\bibitem{BERNA} R. Bernabei et al., INFN/AE-98/34, (1998);
  {it Phys. Lett.} {\bf B 389}, 757 (1996);
  {\it Phys. Lett.} {\bf B 424}, 195 (1998);
{\bf B 450}, 448 (1999).
\bibitem{EDELWEISS}
A. Benoit {\it et al}, [EDELWEISS collaboration], Phys. Lett. B
{\bf 545}, 43 (2002); V. Sanglar,[EDELWEISS collaboration];
D.S. Akerib {\it et al},[CDMS Collaboration], Phys. Rev D {\bf
68}, 082002 (2003);arXiv:astro-ph/0405033.
\bibitem{ref1}For more references see e.g. our previous report:\\
J.D. Vergados,  Supersymmetric Dark Matter Detection- The
Directional Rate and the Modulation Effect, hep-ph/0010151;
\bibitem{ref2}
A. Bottino {\it et al.}, {\it Phys. Lett B} {\bf  402}, 113 (1997).
R. Arnowitt. and P. Nath, {\it Phys. Rev. Lett.}  {\bf 74}, 4952
(1995);
 {\it Phys. Rev. D} {\bf 54}, 2394 (1996); hep-ph/9902237;
 V.A. Bednyakov,  H.V. Klapdor-Kleingrothaus and  S.G. Kovalenko,
{\it Phys. Lett. B}  {\bf  329}, 5 (1994).
\bibitem{Gomez}  M.E.G\'{o}mez and  J.D. Vergados, {\it Phys. Lett. B} {\bf 512}
, 252 (2001); hep-ph/0012020.
 M.E. G\'{o}mez,  G. Lazarides and C. Pallis, Phys. Rev. D {\bf 61}, 123512 (2000) and
 {\it Phys. Lett. B} {\bf 487}, 313 (2000);
 M.E. G\'{o}mez and J.D. Vergados, hep-ph/0105115.
\bibitem{JDV96} 
J.D. Vergados, {\it J. of Phys. G} {\bf  22}, 253 (1996);    
T.S. Kosmas and  J.D. Vergados, {\it Phys. Rev. D} {\bf 55},
1752 (1997).
\bibitem{ARNDU}
 A. Arnowitt and B. Dutta, Supersymmetry and Dark Matter,
hep-ph/0204187;.
 E. Accomando, A. Arnowitt and B. Dutta,
hep-ph/0211417.
\bibitem{KVprd}T.S. Kosmas and  J.D. Vergados, {\it Phys. Rev. D} {\bf 55},
1752 (1997).
\bibitem{VQS03} J.D. Vergados, P.Quentin and D. Strottman,
 (to be published)
\bibitem{Ress} M.T. Ressell {\it et al.}, {\it Phys. Rev. D} {\bf  48},
 5519 (1993);
M.T. Ressell and D.J. Dean, Phys. Rev. C {\bf 56} (1997) 535.
\bibitem{DIVA00} P.C. Divari, T.S. Kosmas, J.D. Vergados and L.D. Skouras,
{\it  Phys. Rev. C} {\bf 61} (2000), 044612-1.
\bibitem{DFS86}
A.K. Drukier, K. Freese and D.N. Spergel, {\it Phys. Rev. D} {\bf
33}, 3495 (1986);
K. Frese, J.A Friedman, and A. Gould, {\it Phys. Rev. D} {\bf 37},
3388 (1988).
\bibitem{Verg}
J.D. Vergados, {\it Phys. Rev. D} {\bf  58}, 103001-1 (1998);
 {\it Phys. Rev. Lett} {\bf  83}, 3597 (1999);
 {\it Phys. Rev. D} {\bf  62}, 023519 (2000);
 {\it Phys. Rev. D} {\bf 63}, 06351 (2001).
 \bibitem{SUHONEN03}
E. Homlund, M. Kortelainen, T.S. Kosmas, J. Suhonen and J.
Toivanen, Phys. Lett B {\bf 584} (2004) 31;\\
 Phys. Atom. Nucl. {\bf 67} (2004) 1198 .
\bibitem{COLLAR92}
J.I. Collar {\it et al}, {\it Phys. Lett. B} {\bf 275}, 181
(1992);
P. Ullio and M. Kamioknowski, {\it JHEP} {\bf 0103}, 049 (2001);
P. Belli, R. Cerulli, N. Fornego and S. Scopel, {\t Phys. Rev. D}
{\bf 66}, 043503 (2002); 
A. Green, {\it Phys. Rev. D} {\bf 66}, 083003 (2002).
\bibitem{GREEN04}
B. Morgan, A.M. Green and N. Spooner, astro-ph/0408047.
\bibitem{DRIFT}
D.P. Snowden-Ifft, C.C. Martoff and J.M.Burwell, Phys. Rev. D {\bf 61}, 1 (2000);\\
M. Robinson {\it et al}, Nucl.Instr. Meth. A {\bf 511}, 347 (2003) 
\bibitem{grypeos}
 S.E.Massen, V.P.Garistov and M.E.Grypeos, Nucl. Phys. A {\bf 597}, 19  19.\\
 A.N. Antonov, S.S. Dimitrova, M.K. Gaidarov, M.V. Stoitsov, M.E. Grypeos,
 S.E. Massen and K.N. Ypsilantis, Nucl.Phys. A {\bf 597}, 163 (1996) \\
C. Moustakidis, S.E. Massen, C.P. Panos, M.E. Grypeos and  A.N. Antonov,
 Phys. Rev. C {\bf 64}, 014314 (2001)
 \bibitem{JDV03}
J.D. Vergados, Phys. Rev. D {\bf 67} (2003) 103003; {\it ibid}
{\bf 58} (1998) 10301-1;
 J.D. Vergados, J. Phys. G: Nucl. Part. Phys. {\bf 30},
1127 (2004).
\bibitem{idm04}
J.D. Vergados, hep-ph/0410378, to appear in the idm2004 proceedings
\bibitem{VEJ04}
J.D. Vergados and H. Ejiri, hep-ph/0401158 (to be published)
\bibitem{GV03}
Y. Giomataris and J.D. Vergados, Neutrinos in a spherical box, to
appear in Nucl.Instr. Meth. (to be published) $\&$
hep-ex/0303045.
\end{thebibliography}
\end{document}